\documentclass[aps,prd,twocolumn,preprintnumbers,floatfix,nofootinbib]{revtex4-1}
\pdfoutput=1
\usepackage{graphicx}
\usepackage{bm}
\usepackage{times}
\usepackage{slashed}
\usepackage[usenames,dvipsnames,svgnames,table]{xcolor}
\usepackage[normalem]{ulem}
\usepackage{lipsum}
\usepackage{subfigure}
\usepackage{amsfonts}		
\usepackage{amsmath}		
\usepackage{amssymb}		
\newcommand{\be}{\begin{equation}}
\newcommand{\ee}{\end{equation}}
\newcommand{\bea}{\begin{eqnarray}}
\newcommand{\eea}{\end{eqnarray}}

\newcommand{\HS}{\texttt{HiggsSignals}}
\newcommand{\HSv}[1]{\texttt{HiggsSignals-#1}}

\newcommand{\HBv}[1]{\texttt{HiggsBounds-#1}}

\newcommand{\FHv}[1]{\texttt{FeynHiggs-#1}}

\newcommand{\MeV}{~\mathrm{MeV}}
\newcommand{\gev}{~\mathrm{GeV}}
\newcommand{\tev}{~\mathrm{TeV}}

\newcommand{\cp}{\mathcal{CP}}
\newcommand{\CL}[1]{#1\%~\mathrm{C.L.}}
\newcommand{\order}[1]{\mathcal{O}(#1)}

\newcommand{\neut}[1]{\tilde{\chi}_#1^0}

\newcommand{\mneut}[1]{m_{\tilde{\chi}_#1^0}}

\newcommand{\mstop}[1]{m_{\tilde{t}_#1}}

\begin{document}

\title{Alignment without Decoupling: the Portal to Light Dark Matter in the MSSM}

\author{Stefano Profumo$^1$}
\email{profumo@ucsc.edu}
\author{Tim Stefaniak$^1$}
\email{tistefan@ucsc.edu}

\preprint{SCIPP 16/12}

\affiliation{$^1$Department of Physics and Santa Cruz Institute for Particle Physics, University of California, Santa Cruz, CA 95064, USA}

\begin{abstract}
We study light, thermal neutralino dark matter in the sub-$\mathrm{GeV}$ to $\sim 65\gev$ mass range in the Minimal Supersymmetric extension of the Standard Model (MSSM). We consider realizations of the limit of alignment without decoupling in the Higgs sector where the heavier $\cp$-even Higgs impersonates the observed Higgs state at $125\gev$, while the lighter $\cp$-even Higgs is the mediator of dark matter annihilation.
We single out three distinct and novel possibilities for light dark matter: (\emph{i}) a neutralino with mass around half the light Higgs mass, in the sub-$\mathrm{GeV}$ to $\sim 30\gev$ mass range; (\emph{ii}) a neutralino with a mass around half the pseudoscalar Higgs boson mass, in our examples  around $(60 - 65)\gev$; (\emph{iii}) a very light neutralino with mass around the light Higgs mass, pair-annihilating to Higgs pairs. We discuss the implications of all these possibilities for indirect and direct dark matter detection experiments, and we demonstrate that all scenarios will be tested by next generation direct detection experiments.
We also emphasize that the unique Higgs phenomenology of this setup warrants a dedicated search program at the LHC.

\end{abstract}

\maketitle

\section{Introduction}
Softly-broken electroweak-scale supersymmetry has long been considered a compelling solution to both the hierarchy and the dark matter problems \cite{Jungman:1995df, Drees:2004jm, Dine:2007zp}. Null results from searches for supersymmetric particles as well as the discovery of a Standard Model (SM) Higgs-like scalar state with a mass around $125\gev$~\cite{ATLASdiscovery,CMSdiscovery} at the Large Hadron Collider (LHC) have excluded a number of possible incarnations of low-energy supersymmetry, and have greatly constrained the range of realizations for supersymmetric dark matter. In particular, within the Minimal Supersymmetric extension of the Standard Model (MSSM)~\cite{Nilles:1983ge,Haber:1984rc,Barbieri:1987xf}, it has been argued~\cite{Boehm:2013gst,Calibbi:2013poa, Cahill-Rowley:2014twa, Calibbi:2014lga} that thermal neutralino dark matter lighter than $30\gev$ is no longer a possibility, given the Higgs mass measurements, constraints from the invisible decay modes of the $Z$ boson, and mass limits for charged supersymmetric particles from the Large Electron-Positron Collider (LEP).

The Higgs sector of the $\cp$-conserving MSSM consists of two Higgs doublets, $H_u$ and $H_d$, that give rise to five physical Higgs states after electroweak symmetry breaking (EWSB): two $\cp$-even neutral Higgs bosons, $h$ and $H$ (where $h$ is lighter than $H$ by definition), one $\cp$-odd neutral Higgs boson, $A$, and one charged Higgs state (and its conjugate), $H^\pm$. At tree-level, the MSSM Higgs sector can be described by only two parameters, often chosen to be the $\cp$-odd Higgs mass, $M_A$ (or alternatively the charged Higgs mass, $M_{H^\pm}$), and $\tan\beta\equiv v_d/v_u$, where $v_{u,d}$ are the vacuum expectation values of the neutral, real components of $H_{u,d}$. 

In order to be consistent with LHC data~\cite{ATLASdiscovery,CMSdiscovery}, the MSSM must yield  a mass value of $\sim 125\gev$ for one of the $\cp$-even MSSM Higgs bosons, and signal rates close enough to those predicted for the SM Higgs boson. Such a \emph{SM-like} Higgs boson can be obtained in two different ways: 

(\emph{i}) the light $\cp$-even Higgs boson $h$ obtains SM-like couplings in the \emph{decoupling limit}, where the remaining Higgs states are rather heavy ($m \gtrsim 300\gev$)~\cite{Haber:1989xc,Haber:1995be,Dobado:2000pw,Gunion:2002zf}; 

(\emph{ii}) the light Higgs boson $h$ \emph{or} the heavy Higgs boson $H$ obtains SM-like couplings in the limit of \emph{alignment without decoupling}~\cite{Carena:2013ooa,Carena:2014nza,Dev:2014yca}, obtained for rather specific parameter space choices (see below).

Ref.~\cite{Bechtle:2016kui} recently demonstrated that both scenarios are viable in light of the combined constraints from Higgs mass and rate measurements, LHC searches for non-SM-like Higgs bosons and supersymmetric particles as well as low energy observables including the rare $B$ decays $b\to s\gamma$, $B_s\to \mu^+\mu^-$ and $B_u \to \tau\nu$ (see also Refs.~\cite{Bechtle:2012jw,Scopel:2013bba,Djouadi:2013lra,Cheung:2015uia,deVries:2015hva,Henrot-Versille:2013yma,Bertone:2015tza,Bhattacherjee:2015sga,Barman:2016jov} and~\cite{Heinemeyer:2011aa,Hagiwara:2012mga,Benbrik:2012rm,Drees:2012fb,Han:2013mga,Bechtle:2012jw,Bhattacherjee:2013vga} for related studies of $h$ or $H$, respectively, being the MSSM candidate for the scalar discovered at the LHC.) In particular, both the light and the heavy Higgs interpretation of the $125\gev$ scalar state achieved in the limit of alignment without decoupling are viable.

The MSSM scenario of alignment without decoupling in the Higgs sector --- in both its incarnations with a SM-like light or heavy Higgs boson --- is an exciting possibility for the LHC, as the remaining Higgs states are expected to be light and within discovery reach in the near future. In the MSSM, this limit can only occur through an accidental cancellation of tree-level and higher-order contributions to the $\cp$-even Higgs mass matrix squared. In turn, this only happens for a large Higgsino mass parameter, $\mu$, and (in some cases) a large trilinear coupling for the scalar top (stop) quark, $A_t$.\footnote{The approximate one-loop contributions relevant for the MSSM alignment without decoupling scenario have been discussed in Refs.~\cite{Carena:2013ooa,Carena:2014nza}. Leading two-loop corrections have been briefly addressed in Ref.~\cite{Bechtle:2016kui} and will be assessed in more detail in an upcoming publication~\cite{preparation}.} More precisely, in order for alignment without decoupling to occur at reasonably low $\tan\beta$ values that are experimentally allowed one needs $\mu/M_S \gtrsim \order{2-3}$~\cite{Bechtle:2016kui}, where the SUSY mass scale $M_S$ is defined as the geometric mean of the two stop masses, $M_S \equiv \sqrt{\mstop{1}\mstop{2}}$.

In this work we explore the dark matter (DM) phenomenology of the alignment without decoupling scenario of the MSSM Higgs sector, and we consider the lightest neutralino, $\neut{1}$, as the DM candidate. In the MSSM the lightest neutralino is a linear combination of the fermionic superpartners of the $U(1)_Y$ and neutral $SU(2)_L$ gauge bosons --- the bino, $\tilde{B}$, and wino, $\tilde{W}$, respectively --- and the two neutral Higgs superpartners, the so-called Higgsinos, $\tilde{h}_{1,2}$. However, in the limit of alignment without decoupling, the Higgsino components are very small because $\mu$ needs to be large\footnote{Barring the possibility of extremely large lightest neutralino masses~\cite{Profumo:2005xd}.}.

Several possibilities exist within the MSSM for the lightest neutralino to be a viable thermal relic with an abundance matching the inferred DM abundance from cosmological observations~\cite{Ade:2015xua}, including e.g.~coannihilation with sfermions, or a well-tempered mixture of its wino and bino components (see e.g.~Ref.~\cite{Baer:2005jq}). These mechanisms are generally also possible in the alignment without decoupling scenario.
Here, however, we focus on the possibility that supersymmetric dark matter is light, i.e.~with mass well below $100\gev$, and is realized via mechanisms that can only arise in the limit of alignment without decoupling. Indeed, we show that in the case of a SM-like heavy $\cp$-even Higgs boson, a window for {\em very light} (even sub-GeV) supersymmetric dark matter is possible.

There are various reasons that motivate exploring this possibility. One is the current interest on light dark matter searches with direct detection experiments (see e.g.~Ref.~\cite{Cushman:2013zza}); another is that a rather interesting collider phenomenology may arise due to the interplay of a new light stable neutral particle and light non-SM Higgs bosons. Finally, a light Higgs boson that mediates DM annihilation pushes the boundaries of how light the cold thermal neutralino relic can be. Traditionally, the lower neutralino mass limit is associated with Lee-Weinberg type limits~\cite{leeweinberg} at around $(5-10)\gev$ because of the lack of a light mediator in the MSSM (see e.g.~Ref.~\cite{Dreiner:2009ic}).


In the remainder of this work we will focus on the case where the heavier of the two $\cp$-even Higgs bosons impersonates the SM-like Higgs particle observed at the LHC. This rather unusual setup can be achieved only in a parameter region subject to various experimental constraints~\cite{Bechtle:2016kui}. We revisit here the constraints arising from Higgs boson searches at LEP and LHC, and comment on other constraints from radiative Upsilon decays which are relevant at very low masses of the light $\cp$-even Higgs boson. Furthermore, and going beyond the discussion in Ref.~\cite{Bechtle:2016kui}, we demonstrate numerically that the decay $H\to hh$ vanishes in the limit of alignment without decoupling, while the decay $A\to Zh$ is unsuppressed and provides an interesting new collider signature. 

We then select four benchmark points that are compatible with all collider constraints in order to illustrate that the lightest neutralino can yield the correct thermal relic density either via  resonant $s$-channel light Higgs or pseudo-scalar Higgs exchange, for neutralino masses around half the Higgs mass, or via annihilation to a final state consisting of a pair of light Higgs, for comparable neutralino and light Higgs masses.

The outline of this paper is as follows: Section~\ref{sec:benchmarks} describes our choice for the MSSM parameters that give rise to the limit of alignment without decoupling with the heavy $\cp$-even Higgs boson as the observed SM-like Higgs state at $125\gev$. We give details on the Higgs phenomenology and experimental constraints in this scenario, and select four illustrative benchmark points. Section~\ref{sec:dm} focuses on the phenomenology of the lightest neutralino as a thermal relic dark matter candidate, illustrates the mechanisms for efficient dark matter annihilation, and discusses the implications for direct and indirect dark matter searches. We present a final discussion of our results and our conclusions in Section~\ref{sec:conclusions}.

\section{Higgs Phenomenology and Benchmarks}
\label{sec:benchmarks}

\begin{figure*}
\centering
\includegraphics[width=0.48\textwidth]{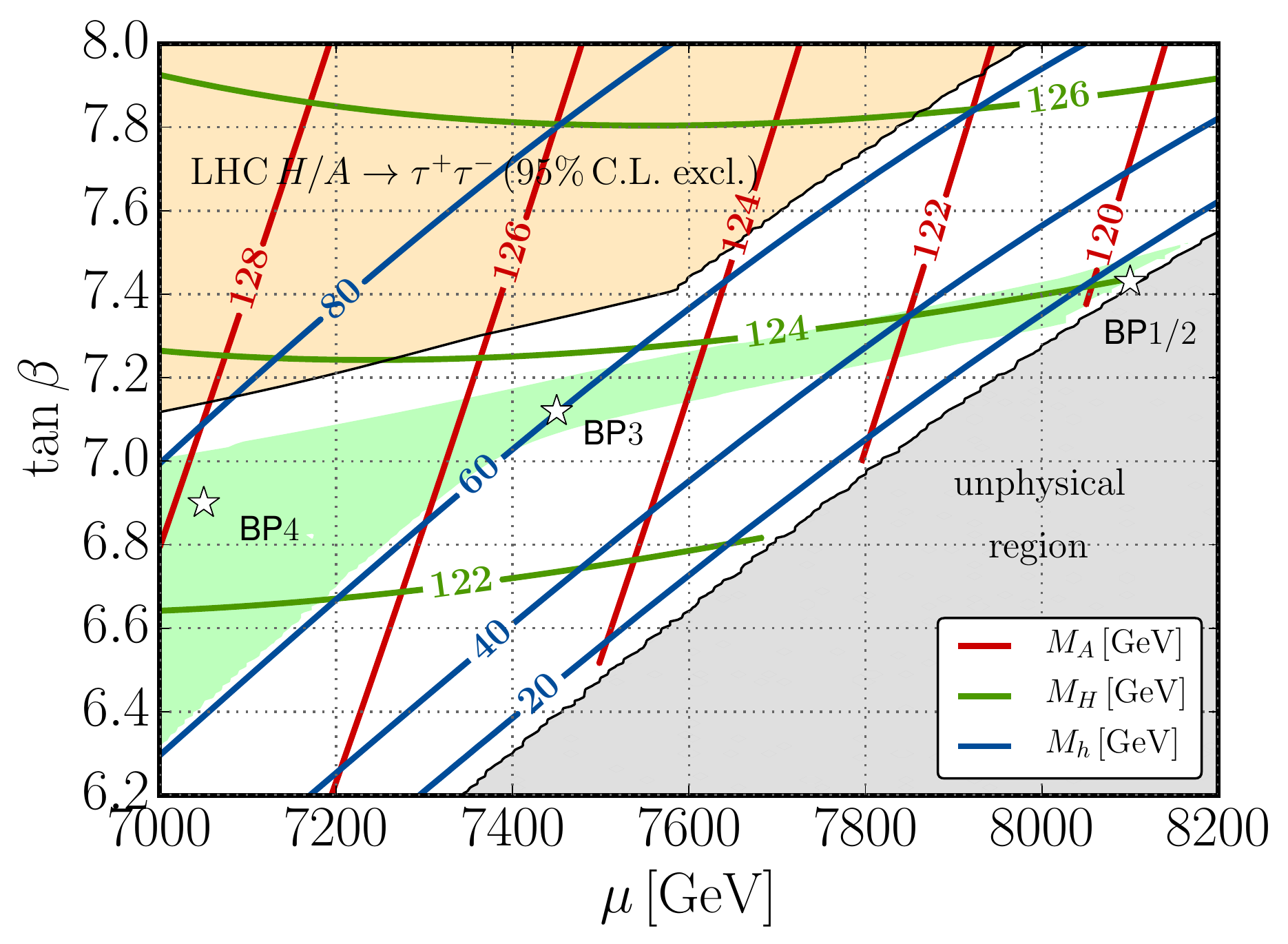}\hfill
\includegraphics[width=0.48\textwidth]{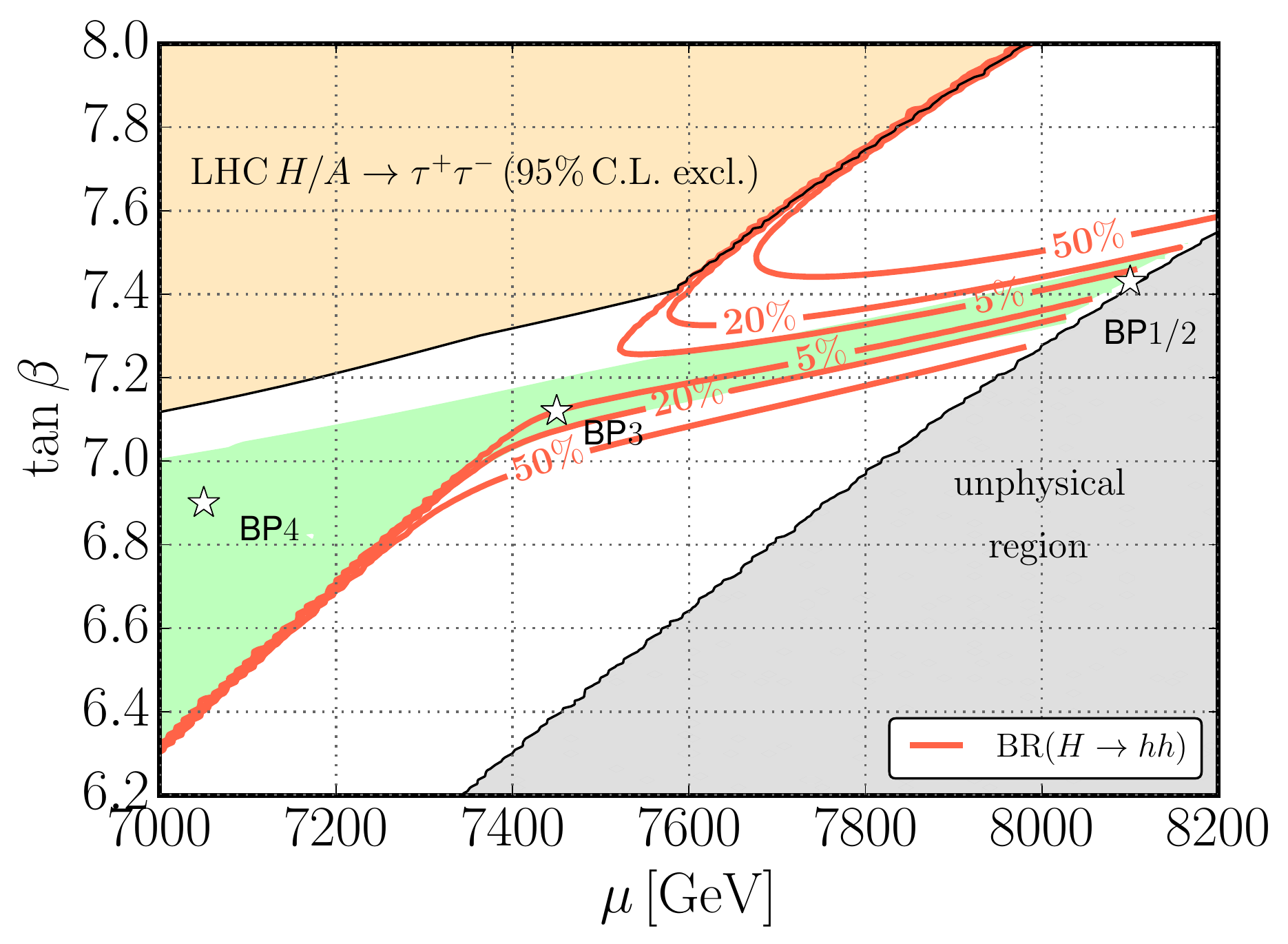}
\caption{Representative ($\mu$, $\tan\beta$) parameter region for the heavy Higgs interpretation of the LHC Higgs signal. The parameters are chosen according to Tab.~\ref{Tab:pars}. 
The green region indicates good agreement with the Higgs signal rates (see text), the orange region is excluded (at $\CL{95}$) by LHC $H/A\to \tau^+\tau^-$ searches and the gray area shows an unphysical region with negative Higgs mass. The white stars locate the selected benchmark points, see Tab.~\ref{Tab:BPs}. \emph{Left}: contours give the neutral Higgs boson masses; \emph{Right}: contours indicate the Higgs-to-Higgs branching fraction $\text{BR}(H\to hh)$.}
\label{Fig:BP}
\end{figure*}

In this section we discuss the parameter choices that lead to the limit of alignment without decoupling with the {\em heavy} $\cp$-even Higgs boson playing the role of the $125\gev$ scalar particle observed at the LHC, and the associated phenomenology of the Higgs sector. At the end of this section we select four representative benchmark points that will be employed in Section~\ref{sec:dm} to illustrate the dark matter phenomenology of this scenario.


\begin{table}
\centering
\setlength{\tabcolsep}{10pt}
\renewcommand{\arraystretch}{1.3}
\begin{tabular}{l | l}
\toprule
Parameter & Value \\
\colrule
$M_{H^\pm}$ & $155\gev$ \\
$M_{\tilde{Q}_{1,2}}, M_{\tilde{u}_{1,2}}, M_{\tilde{d}_{1,2}}$ & $2\tev$ \\
$M_{\tilde{Q}_3}, M_{\tilde{u}_3}, M_{\tilde{d}_3}$ & $1\tev$ \\
$M_{\tilde{L}_{1,2}}, M_{\tilde{E}_{1,2}}$ & $250\gev$\\
$M_{\tilde{L}_{3}}, M_{\tilde{E}_{3}}$ & $500\gev$\\
$A_t, A_b, A_\tau$ & $-100\gev$ \\
$M_1$ & $63\gev$\\
$M_2$ & $300\gev$\\
$M_3$ & $2\tev$\\
\botrule 
\end{tabular}
\caption{Choice of the relevant SUSY parameters for the $(\mu, \tan\beta)$ plane shown in Fig.~\ref{Fig:BP} which represents the heavy Higgs interpretation. All parameters are defined in the on-shell (OS) renormalization scheme. See text for the parameter definitions.}
\label{Tab:pars}
\end{table}

Fig.~\ref{Fig:BP} illustrates a typical region in the ($\mu$,$\tan\beta$) parameter plane where the heavy Higgs interpretation can be successfully realized (inspired by the benchmark scenarios presented in Ref.~\cite{Bechtle:2016kui}).\footnote{The heavy Higgs interpretation can only appear at very large $\mu$ values in the MSSM, which exacerbates the ``little hierarchy problem''~\cite{Barbieri:1987fn}, i.e.~the fine-tuning of the squared soft-breaking Higgs mass parameters $M_{H_u}^2$, $M_{H_d}^2$ against $\mu^2$ to achieve a number $\sim M_Z^2$. Nevertheless, if such a fine-tuning in the electroweak sector is accepted, these scenarios are theoretical fully consistent and viable.} Our choice for the remaining SUSY parameters relevant for this scenario is given in Tab.~\ref{Tab:pars} (notice that all parameters are defined in the on-shell (OS) renormalization scheme).
$A_t$, $A_b$ and $A_\tau$ are the soft-breaking trilinear scalar couplings for the top, bottom and tau sector, $M_{\tilde{Q}_i}$, [$M_{\tilde{u}_i}$ and $M_{\tilde{d}_i}$] $(i=1,2,3)$ are the soft-breaking mass parameters for the left-handed [right-handed] $i$th generation [up- and down-type] scalar quarks (squarks), and $M_1$, $M_2$ and $M_3$ are the soft-breaking bino, wino and gluino mass parameters, respectively. Notice that the parameter $M_1$, which has a very minor impact on the Higgs sector, will be varied later in Section~\ref{sec:dm} to study the dark matter phenomenology of each benchmark model.\footnote{In principle, if $M_1$ (and thus the lightest neutralino mass) is small enough, the decay mode $H\to \neut{1}\neut{1}$ becomes kinematically accessible and can alter the Higgs phenomenology considerably. However, in our scenario the Higgsino mass parameter $\mu$ is very large and hence the $H\neut{1}\neut{1}$ coupling is very small. Therefore, the decay rate for $H\to \neut{1}\neut{1}$ is negligible here.}

The mass parameters for the first and second generation squarks and the gluino are set to $2\tev$ to evade the current limits from the LHC. For simplicity, we choose for the left-handed [right-handed] soft-breaking slepton mass parameters of the first/second generation, $M_{\tilde{L}_{1,2}}$ [$M_{\tilde{E}_{1,2}}$], and third generation, $M_{\tilde{L}_{3}}$ [$M_{\tilde{E}_{3}}$], values of $250\gev$ and $500\gev$, respectively,
however the exact values are secondary for the phenomenology discussed here.\footnote{A light stau can significantly enhance the Higgs decay width to diphotons, $\Gamma(H\to \gamma\gamma)$ (see e.g.~Refs.~\cite{Djouadi:2005gj,Carena:2012gp,Liebler:2015ddv}). Indeed, lowering the stau mass in our benchmark scenarios would slightly improve agreement with the Higgs data. Additionally, a light smuon would produce a SUSY contribution to the anomalous magnetic moment of the muon that would bring closer agreement with experimental data~\cite{Bechtle:2016kui}.} We set the top quark pole mass to $m_t = 173.2\gev$~\cite{Agashe:2014kda}.

We employ \FHv{2.12.0}~\cite{Heinemeyer:1998np,Heinemeyer:1998yj,Degrassi:2002fi,Frank:2006yh,Hahn:2009zz,Hahn:2013ria} to evaluate the SUSY and Higgs mass spectrum as well as the Higgs production cross sections and decay rates. We test limits from Higgs searches at LEP and LHC using \HBv{4.3.1}~\cite{Bechtle:2008jh,Bechtle:2011sb,Bechtle:2013gu,Bechtle:2013wla,Bechtle:2015pma}. In particular, we obtain the limits from the CMS $H/A\to \tau^+\tau^-$ search at $7/8\tev$~\cite{Khachatryan:2014wca,CMS:2015mca} using the likelihood implementation described in Ref.~\cite{Bechtle:2015pma}.\footnote{For low $M_A$ values (as discussed here) the $H/A\to \tau^+\tau^-$ search results from the combined $7$ and $8\tev$ run are still more sensitive than the preliminary $13\tev$ results~\cite{CMS:2016pkt}.} The parameter region excluded at $\CL{95}$ from this search is shown in orange in Fig.~\ref{Fig:BP}. We use \HSv{1.4.0}~\cite{Bechtle:2013xfa,Bechtle:2014ewa} to evaluate the $\chi^2$ compatibility with the measured Higgs signal rates from the $7/8\tev$ run, using the included $85$ measurements. The parameter regions that are roughly in agreement with these measurements are loosely defined by $\chi^2/N_{\rm obs} \le 1$, with the number of observables $N_{\rm obs} = 85$. Such regions are shown in green in Fig.~\ref{Fig:BP}. The region of alignment without decoupling corresponds to  $\tan\beta$ values around $7.2$ (depending on $\mu$), where the heavy Higgs boson $H$ becomes SM-like and good agreement with the Higgs data is achieved.

The contour lines in Fig.~\ref{Fig:BP} (\emph{left}) indicate the masses of the neutral Higgs bosons $h$, $H$ and $A$. The heavy $\cp$-even Higgs mass is compatible with the observed Higgs mass $\sim 125\gev$ over most of the parameter region (taking into account a theoretical uncertainty $\Delta M_H\sim \order{2-3}\gev$). In addition, the pseudoscalar Higgs mass, $M_A$, is also in the vicinity of this value, leading to a non-trivial overlap of $H$ and $A$ signals in the experimental analyses of the Higgs signal at the LHC. While $A$ does not couple to gauge bosons at tree-level, leaving the rates of the well-measured Higgs channels with $\gamma\gamma$, $ZZ^*$ and $WW^*$ final states essentially unaffected, its couplings to down-type fermions are enhanced at large $\tan\beta$. Therefore, in these scenarios the rates for the Higgs channels with $b\bar{b}$ and $\tau^+\tau^-$ final states are typically predicted to be slightly larger than for a purely SM-like Higgs boson.\footnote{The Higgs channels with $b\bar{b}$ and $\tau^+\tau^-$ final states have a relatively poor mass resolution, such that the two Higgs bosons $H$ and $A$ would not be seen separately. Note also, that these searches often require Higgs production in association with a $W$ or $Z$ boson, or instead in the process of vector-boson fusion (VBF). These production processes require a Higgs coupling to vector bosons and hence the $A$ would not signify in these searches.} 

In the parameter space of Fig.~\ref{Fig:BP} the light Higgs boson $h$ attains mass values between $0\gev$ and $\sim 80\gev$. At low $M_h \lesssim 60\gev$ the decay $H\to hh$ becomes kinematically accessible. The corresponding decay width is generally rather large, leading to a substantial branching ratio $\text{BR}(H\to hh)$ in most of the parameter region with $M_h < 60\gev$, as shown in Fig.~\ref{Fig:BP} (\emph{right}). However, a notable exception is the  alignment without decoupling region, where the decay $H\to hh$ is highly suppressed. Therefore, in this region the signal rates for the heavy $\cp$-even Higgs boson $H$ remain SM-like even for very low $M_h$. 

Another interesting decay mode appearing at low $M_h$ is the pseudoscalar Higgs decay $A\to Zh$. In contrast to $H\to hh$, this decay mode is unsuppressed in the limit of alignment without decoupling. Moreover, if sizable, the $A\to Zh$ decay mode suppresses the branching ratios for the decays $A\to b\bar{b}$ and $A\to\tau^+\tau^-$, thus leading to a more SM-like signal expectation for experimental analyses of the $125\gev$ Higgs boson in these final states.

Higgs searches at LEP have targeted the light Higgs mass range~\cite{Schael:2006cr}. The main Higgs production channel at LEP is the so-called Higgs-Strahlung process, $e^+e^-\to Zh$, whose cross section depends quadratically on the Higgs-vector boson coupling, $g_{hZZ}\propto \sin(\beta-\alpha)$, where the angle $\beta-\alpha$ diagonalizes the $\cp$-even Higgs squared-mass matrix in the Higgs basis (see e.g.~Ref.~\cite{Carena:2014nza} for details). In the limit of alignment without decoupling where the heavy $\cp$-even Higgs boson is SM-like, we have $\sin(\beta-\alpha) \ll 1$, and thus $h$ production in the Higgs-Strahlungsprocess is highly suppressed. Therefore, the most relevant LEP Higgs searches are those for $e^+e^- \to Ah$, with subsequent decays of the Higgs bosons $A$ and $h$ to $b\bar{b}$ or $\tau^+\tau^-$ pairs. With $A$ masses around $120$ to $130\gev$ and $h$ masses $\lesssim 80\gev$, these processes are kinematically accessible at LEP with center-of-mass energies $\sqrt{s}$ up to $209\gev$. However, the decay $A\to Zh$ becomes sizable at low $M_h$ values and suppresses the expected rate of the above channels. Overall, we find that the limits from these search channels are not strong enough to yield a $\CL{95}$ exclusion in the  parameter region shown in Fig.~\ref{Fig:BP}.

\begin{table*}
\centering
\setlength{\tabcolsep}{10pt}
\renewcommand{\arraystretch}{1.3}
\begin{tabular}{l | cccc}
\toprule
	& BP1	&	BP2	& 	BP3	& 	BP4 \\    
\colrule 
$\mu~[\mathrm{GeV}]$ & $8100$		& $8100$ & 	$7450$	&$7050$	\\
$\tan\beta$ 	&	 $\sim7.42$	&	$7.43$	& $7.12$ & $6.9$\\
\colrule
$M_h~[\mathrm{GeV}]$ &$\sim 0.975$ & $9.8$	&$60.2$ & $75.5$\\
$M_H~[\mathrm{GeV}]$ & $123.9$ & $124.0$  &$123.5$ & $122.8$\\
$M_A~[\mathrm{GeV}]$ & $119.5$ & $119.6$ & $125.2$ &$127.8$ \\
\colrule
$\text{BR}(H\to hh)$ & $0.7\%$ & $0.5\%$  & $4.9\%$ & - \\
$\text{BR}(A\to Zh)$ & $41.2\%$ & $36.6\%$ & $3.6\%$ & $0.6\%$ \\
$\text{BR}(H^+\to \tau^+\nu_\tau)$& $4.0\%$ & $4.1\%$ & $16.5\%$ & $75.3\%$\\
$\text{BR}(H^+\to W^+ h)$ & $96.0\%$ & $95.9\%$ & $83.3\%$ & $23.6\%$\\
\colrule
$\chi^2_\text{HS}/N_\text{obs}$	& $76.4/85$  & $77.4/85$  & $82.0/85$ &  $79.8/85$ \\
\botrule
\end{tabular}
\caption{The four selected benchmark points (BP1-4): values for the input parameters $\mu$ and $\tan\beta$, the neutral Higgs boson mass spectrum, selected Higgs boson decay rates, and the \HS\ chi-squared value, $\chi_{\rm HS}^2$, from the Higgs signal rates over the number of signal rate observables, $N_{\rm obs}$. Parameter values not listed are specified in Tab.~\ref{Tab:pars}. 
For all models $M_1$ is a free parameter that we vary between 0 and $\sim70$ GeV in what follows.}
\label{Tab:BPs}
\end{table*}

For very small $h$ masses, $M_h < M_{\Upsilon(1S)} \simeq 9.46\gev$, radiative Upsilon decays to the light Higgs boson, $\Upsilon(1S) \to h\gamma$, become important~\cite{Wilczek:1977zn}. The \texttt{Babar} experiment at SLAC searched for this process, with $h$ successively decaying to $\mu^+\mu^-$~\cite{Lees:2012iw} and $\tau^+\tau^-$~\cite{Lees:2012te}, and essentially excluded light Higgs masses $M_h$ below around $9\gev$, since the $h$ coupling to down-type fermions is $\tan\beta$ enhanced in our scenario. However, it is noteworthy that small mass gaps exist, where such light Higgs bosons cannot be ruled out. For instance, Higgs masses in the vicinity of the scalar meson $f_0(975)$ resonance cannot be excluded, as the resonant enhancement of the decay $h\to \pi\pi$ suppresses the experimentally observable decay $h\to \mu^+\mu^-$~\cite{Raby:1988qf,Truong:1989my,Donoghue:1990xh} (see also Refs.~\cite{Dawson:1989xs,Gunion:1989we,Clarke:2013aya} for an overview of constraints on very light Higgs bosons). We shall exploit this caveat below to exemplify a successful realization of a sub-$\mathrm{GeV}$ DM candidate in the MSSM.

Current LHC limits from searches for a charged Higgs boson in top quark decays, $t\to H^+b$, with successive charged Higgs decay $H^+\to \tau^+\nu_\tau$, only reach up to $M_{H^\pm} \le  150\gev$~\cite{Aad:2014kga,Khachatryan:2015qxa}. At larger $M_{H^\pm}$ the branching ratio $\text{BR}(t\to H^+b)$ decreases due to phase space. For the parameter plane shown in Fig.~\ref{Fig:BP} we set the charged Higgs mass to $M_{H^\pm} = 155\gev$, see Tab.~\ref{Tab:pars}, 
 such that current charged Higgs limits are evaded. As pointed out in Ref.~\cite{Bechtle:2016kui}, at small light Higgs masses, $M_h\lesssim 70\gev$, the decay $H^\pm \to W^\pm h$ becomes sizable and offers a promising search signature that is complementary to the existing searches for $H^+\to \tau^+\nu_\tau$.

For the ensuing discussion of the DM phenomenology we select four benchmark points (BP) in the parameter plane of Fig.~\ref{Fig:BP} (indicated as stars in the figure). Instead of retaining the bino mass parameter $M_1$ at a fixed value (as e.g.~given in Tab.~\ref{Tab:pars}) 
 we will vary $M_1$ to scan over the lightest neutralino mass, $\mneut{1}$, for each benchmark point.

We list the details of our selected benchmark points in Tab.~\ref{Tab:BPs}, including the chosen $\mu$ and $\tan\beta$ values, the neutral Higgs boson masses, and the branching ratios for the decays $H\to hh$, $A\to Zh$, $H^+\to \tau^+\nu_\tau$ and $H^+\to W^+ h$. We also give the ratio of $\chi^2$ over the number of observables, $N_\text{obs}$, as obtained from \HS\ from the fit to the Higgs signal rates, to indicate that these benchmark points are compatible with the LHC observations of the $125\gev$ Higgs state. 

The parameter values of BP1 and BP2 only differ in the choice of $\tan\beta$, which is used to adjust the mass of the light $\cp$-even Higgs boson to values of $M_h \simeq 975\MeV$ and $9.8\gev$, respectively.\footnote{It should be noted that the theoretical uncertainty of the light Higgs mass is comparable (or even larger) that the predicted value in BP1. Therefore, the exact value of $\tan\beta$ needed for the selected light Higgs mass is unknown. For the sake of reproducibility of our numerical results, the precise value chosen here for BP1 is $\tan\beta =7.424376$.} As mentioned above, the light Higgs mass value of BP1 is allowed due to its proximity to the $f_0(975)$ scalar meson resonance. The pseudoscalar Higgs mass, $M_A \simeq 119.5\gev$, is lower than the heavy $\cp$-even Higgs mass, $M_H \simeq 124.0\gev$. Both benchmark points have sizable branching fractions for the decays $H^+ \to W^+h$ and $A\to Zh$, while $\text{BR}(H\to hh)$ is $\lesssim 1\%$. BP3 features an intermediate light Higgs mass value of $M_h \simeq 60.2\gev$. The masses of the pseudoscalar and heavy $\cp$-even Higgs boson are comparable with $M_A \simeq 125.2\gev$ and $M_H\simeq 123.5\gev$. The branching ratios for the decay $H\to hh$ and the (off-shell) decay $A\to Zh$ are $4.9\%$ and $3.6\%$, respectively. In comparison with the benchmark points BP1 and BP2, this scenario yields a slightly worse fit to the Higgs data, due to the larger contribution of $A$ to the predicted Higgs signal at $\sim 125\gev$ in the $\tau^+\tau^-$ channel. Nevertheless, the $\chi^2$ is still acceptable. The last benchmark point, BP4, has a larger light Higgs mass of $M_h \simeq 75.5\gev$, and thus the decay $H\to hh$ is kinematically closed. The pseudoscalar and the heavy Higgs boson masses are $127.8\gev$ and $122.8\gev$, respectively. The charged Higgs predominantly decays via $H^+\to\tau^+ \nu_\tau$.

\section{Dark matter phenomenology}\label{sec:dm}

\begin{figure*}
\centering
\includegraphics[width=\textwidth]{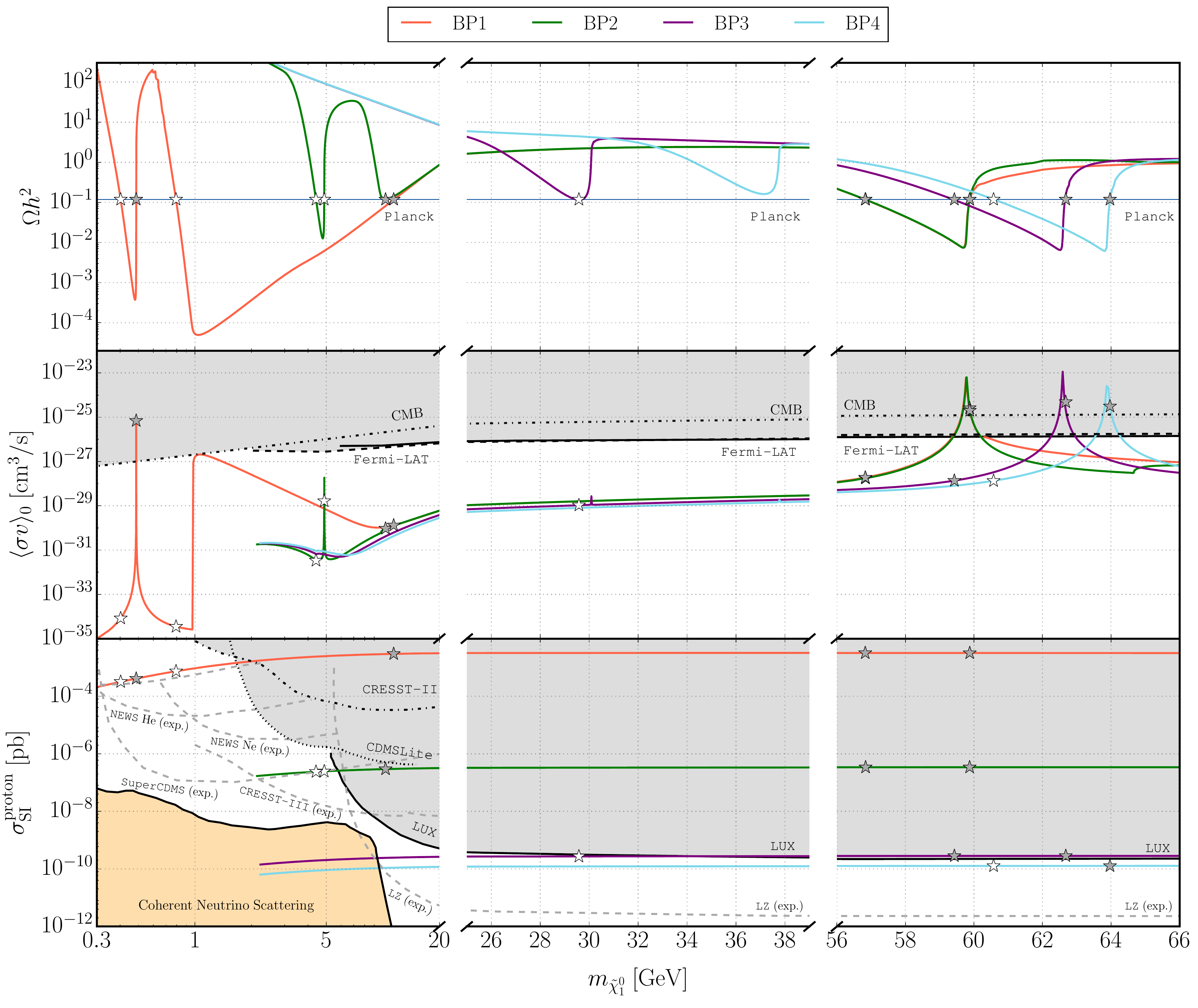}
\caption{Dark matter phenomenology for the four benchmark points, as a function of the lightest neutralino mass, $\mneut{1}$. The white [gray] stars indicate values of $\mneut{1}$ producing a thermal neutralino relic density that matches the observed cosmological dark matter density, and that are allowed [excluded] by direct and [or] indirect detection constraints. \emph{Top panel}: Dark matter thermal relic density, $\Omega h^2$, with the narrow blue line indicating the observed value (including its $2\sigma$ uncertainties in the line width), as determined by the \texttt{Planck} collaboration \cite{Ade:2015xua}; \emph{Center panel}: Thermal average of the annihilation cross section times the relative velocity at zero temperature, $\langle \sigma v\rangle_0$, along with constraints from CMB (dot-dashed)~\cite{Slatyer:2015jla} and \texttt{Fermi-LAT} observations of local dwarf spheroidal galaxies assuming pure $b\bar{b}$ (solid) or $\tau^+\tau^-$ (dashed) final states~\cite{Ackermann:2015zua}; \emph{Bottom panel}: Cross section for spin-independent scattering off of protons, along with the present limits from \texttt{CRESST-II} \cite{Strauss:2016sxp}, \texttt{CDMSLite} \cite{Agnese:2015nto}, \texttt{LUX}, as well as future projected sensitivity curves for \texttt{CRESST-III} \cite{Strauss:2016sxp} and \texttt{LZ} \cite{Akerib:2015cja}, \texttt{NEWS} \cite{Gerbier:2014jwa} and \texttt{SuperCDMS}~\cite{LopezAsamar:2016def}. The orange region indicates the irreducible background from coherent neutrino scattering in direct detection experiments \cite{Cushman:2013zza}.}\label{fig:DM_Mneut1}
\end{figure*}

In this section we explore the light dark matter phenomenology associated with the benchmark points specified above. We vary the single free MSSM parameter of relevance to define the properties of the lightest neutralino as a dark matter candidate,  the soft-supersymmetry breaking bino mass, $M_1$. Because the wino and Higgsino mass parameters, $M_2$ and $\mu$, respectively, are chosen to be much larger (see Tabs.~\ref{Tab:pars} 
and~\ref{Tab:BPs}), this is in practice equivalent, up to small corrections, to varying the lightest neutralino mass, $\mneut{1}$. All quantities related to dark matter have been calculated using the most recent available version (4.3) of the {\tt micrOMEGAs} code \cite{Belanger:2001fz,Belanger:2004yn,Belanger:2014vza}. We fix the target neutralino thermal relic abundance using the \texttt{Planck} results \cite{Ade:2015xua} on the cold dark matter abundance, as inferred from combined cosmological observations, 
\begin{align}\label{eq:planck}
\Omega_{\rm DM} h^2 = 0.1184\pm  0.0012.
\end{align}

As far as indirect dark matter detection is concerned, we calculate the thermally averaged pair-annihilation cross section times relative velocity at zero temperature, $\langle\sigma v\rangle_0$, as a function of the dark matter particle mass, and we consider Cosmic Microwave Background (CMB) limits on ionization from particles injected through dark matter pair-annihilation in the ``dark ages'' \cite{Slatyer:2015jla}. Using the results of Ref.~\cite{Ade:2015xua}, Ref.~\cite{Slatyer:2015jla} finds the limit
$$
f_{\rm eff}\frac{\langle\sigma v\rangle_0}{\mneut{1}}<4.1\times 10^{-28}\ {\rm cm}^{3}{\rm s}^{-1}{\rm GeV}^{-1},
$$
where for the final states relevant in the MSSM the effective ``efficiency factor'' is here $f_{\rm eff}\simeq 0.2$. 
We also consider limits from the non-observation of gamma radiation from local dwarf spheroidal galaxies with the \texttt{Fermi Large Area Telescope (LAT)}~\cite{Ackermann:2015zua}, assuming a $\bar b b$ or $\tau^+\tau^-$ pair-annihilation final state.

Fig.~\ref{fig:DM_Mneut1} presents our key findings on the dark matter phenomenology. We show three dark matter-related quantities as a function of the lightest neutralino mass, $\mneut{1}$, for the four benchmark points detailed upon in Table \ref{Tab:BPs} (the color coding is indicated at the top of the figure). We select three relevant intervals in the lightest neutralino mass to illustrate the salient features. Stars along the slopes for each benchmark scenario indicate neutralino mass values with the correct thermal relic dark matter abundance, Eq.~\eqref{eq:planck}. The white (dark) stars correspond to points which are allowed (excluded) by indirect and (or) direct DM searches.

The top panel of Fig.~\ref{fig:DM_Mneut1}  shows the thermal relic neutralino density, $\Omega h^2$, assuming a standard cosmological history. The horizontal line shows the \texttt{Planck} determination for the cold dark matter abundance, Eq.~\eqref{eq:planck}, and its 2$\sigma$ (barely visible) range. All four benchmark models exhibit a similar pattern in their thermal relic density, which is easily understood:
\begin{enumerate}
\item A first dramatic dip in $\Omega h^2$, driven by resonant annihilation mediated by the light $\cp$-even Higgs boson $h$, at $\mneut{1}\simeq M_h/2$;
\item A second drop in $\Omega h^2$, corresponding to the $hh$ annihilation final state becoming kinematically accessible, at $\mneut{1}\gtrsim M_h$;
\item A fourth and final drop in $\Omega h^2$ associated with resonant pair-annihilation through the pseudoscalar Higgs $A$.
\end{enumerate}
For BP1, BP2 and BP3, the first dip in $\Omega h^2$, corresponding to resonant $s$-channel annihilation via the light $\cp$-even Higgs boson $h$, gives rise to two values of $\mneut{1}$, slightly above and below the resonance at $M_h/2$, where the observed $\Omega h^2$ is produced. In BP3  these two values are essentially degenerate.
Notice that in the $\mneut{1}\lesssim M_h/2$ case, since the center-of-mass energy squared at a finite temperature $T$ is approximately 
\begin{align}
s\simeq 4\mneut{1}^2+6\mneut{1}T,
\end{align}
the $h$ exchange is on-shell for a mass splitting between the lightest neutralino and half the $h$ mass of order the freeze-out temperature, which, in turn, is of the order of $T_{\rm f.o.}\simeq \mneut{1}/(20-25)$.
Conversely, when $\mneut{1}\gtrsim M_h/2$ the finite neutralino kinetic energy brings it more off-shell at finite temperature, hence the sudden drop in the effective pair-annihilation cross section past the natural width of the exchanged scalar. This explains the asymmetric shape of the resonance in Fig.~\ref{fig:DM_Mneut1}.

The second drop mentioned above effectively brings the thermal relic density to low-enough values only for BP1 and BP2, while the third drop, associated with the $s$-channel exchange of the pseudoscalar Higgs boson $A$ is enough for all benchmark points to hit a low-enough thermal relic density.

Notice that the slight difference of BP1 and BP2 for the relic density and indirect detection cross section seen at larger neutralino masses, $\mneut{1}\sim (60-66)\gev$, is due to the opening of the annihilation channel $\neut{1}\neut{1}\to Ah$, which appears at different mass values due to the different light Higgs mass.

The general features of the thermally averaged cross section times velocity $\langle\sigma v\rangle_0$ are understood along similar lines to what we described above, except for one important caveat: at low temperature, since two Majorana fermions in an $s$-wave are in a $\cp$-odd state, the pair-annihilation cross section from $\cp$-even Higgs boson exchange is heavily suppressed. This explains why the resonance associated with the $\cp$-odd scalar in the central right panel of Fig.~\ref{fig:DM_Mneut1} is several orders of magnitude larger than those associated with the $\cp$-even scalar.
As a result, current indirect detection constraints, which would otherwise jeopardize low-mass dark matter models \cite{Slatyer:2015jla}, fail at ruling out the points with very light dark matter masses associated with BP1 and BP2, with only one exception: in the $\mneut{1}\gtrsim M_h/2$ case for BP1 the pair-annihilation cross section, while $p$-wave suppressed, is still large enough to violate the limit from CMB distortions, and it occurs for almost exactly on-shell $h$ exchange. In all other cases (apart from those close to the $A$ resonance), the neutralino masses corresponding to the right thermal relic density yield $\langle\sigma v\rangle_0$ values well below current constraints from either CMB or from gamma-ray observations, below or at most around $10^{-29}\ {\rm cm}^3{\rm s}^{-1}$. 

For reference, we list in Tab.~\ref{Tab:inddir} all lightest neutralino mass values, for each benchmark scenario, which give the right thermal relic dark matter density. The third column gives the dominant pair-annihilation final state at zero temperature, and the last column summarizes whether or not the given mass value is allowed by other constraints (and specifying which constraint rules it out, if that is the case). The table illustrates that for masses between 4 and 30 GeV the dominant annihilation final state is $\tau^+\tau^-$. The corresponding constraints on the pair-annihilation cross section into $\tau^+\tau^-$ final states from \texttt{Fermi-LAT} observations of local dwarf spheroidal galaxies are shown in Fig.~\ref{fig:DM_Mneut1} as a dashed line in the middle panel.

Around the pseudoscalar Higgs resonance indirect detection constraints are {\em not} velocity-suppressed, and in fact they rule out, in all cases, points on the $\mneut{1}\gtrsim M_A/2$ side of the resonance. Due to the same finite-temperature effects mentioned above, the neutralino mass on the $\mneut{1}\gtrsim M_A/2$ side yielding the correct thermal relic density is much closer to on-shell at zero temperature than the one on the $\mneut{1}\lesssim M_A/2$ side, yielding a much larger  $\langle\sigma v\rangle_0$ that is firmly excluded both by CMB and gamma-ray constraints for all four benchmarks.

The bottom panels show our results for direct dark matter searches. We show with a black curve the most recent (July 2016) limits from the \texttt{LUX} Collaboration~\cite{LUX}, while the dashed curves indicate the projected sensitivity of the future \texttt{LZ} experiment~\cite{Akerib:2015cja}. In the low-mass region, we show the 2015 results from the \texttt{CRESST-II} Collaboration~\cite{Strauss:2016sxp} and from \texttt{CDMSLite}~\cite{Agnese:2015nto}. We also indicate the future sensitivity of the \texttt{CRESST-III} Phase 2 detector and of the dedicated low-mass dark matter direct search experiments \texttt{NEWS}~\cite{Gerbier:2014jwa} and \texttt{SuperCDMS}~\cite{LopezAsamar:2016def}. We also include the coherent neutrino background scattering limit~\cite{Cushman:2013zza}. 

The striking scaling of the direct detection cross section for the four benchmark models under consideration, shown in the three lower panels, is straightforward to understand: The scattering is dominated by $t$-channel exchange of the light Higgs boson $h$, and thus the cross section scales as $M_h^{-4}$.  As a result, for instance, points with the correct relic density at large neutralino masses for the benchmark scenarios BP1 and BP2 are ruled out by the recent LUX results. Only at very low neutralino masses, $\mneut{1}\lesssim 5\gev$, current direct detection experiments are unable to exclude the viable points for BP1 and BP2. For BP3 we find the candidate point(s) at the light Higgs resonance, $\mneut{1}\lesssim M_h/2\simeq 30\gev$, to be right at the edge of the LUX observed limit, while the points around the $A$ resonance are excluded.
 
Rather interestingly, comparing the predicted spin-independent scattering cross section for BP1 and BP2 with the planned sensitivity of, for instance, the \texttt{NEWS} detector with a He target \cite{Gerbier:2014jwa} or the \texttt{CRESST-III} detector~\cite{Strauss:2016sxp} we find that in all cases future experiments will probe light (i.e.~sub-$\mathrm{GeV}$ to $5\gev$) supersymmetric dark matter in the present context.

\begin{table*}
\centering
\renewcommand{\arraystretch}{1.3}
\begin{tabular}{ c |c | ccc | c}
\toprule
Benchmark scenario 	& $m_{\tilde{\chi}^0_1}$~[GeV] & \multicolumn{3}{c|}{dominant pair annihilation final states}  & allowed? (relevant constraint)\\
\colrule
BP1		&	$0.40$	&	$gg~(55.6\%)$ &	$s\bar{s}~(20.9\%)$ & $d\bar{d}~(20.9\%)$ & yes\\
BP1		&	$0.49$	&	$gg~(66.6\%)$ &	$s\bar{s}~(16.7\%)$ & $d\bar{d}~(16.7\%)$ & no (CMB)\\
BP1		&	$0.79$	&	$gg~(74.4\%)$ &	$s\bar{s}~(8.1\%)$ & $d\bar{d}~(8.1\%)$ & yes\\
BP2		&	$4.4$	&	$\tau\tau~(98.5\%)$ & $c\bar{c}~(1.1\%)$ &  & yes\\
BP2		&	$4.9$	&	$\tau\tau~(53.1\%)$ &	$b\bar{b}~(41.4\%)$ & $gg~(5.3\%)$ & yes\\
BP1		&	$11.4$	&	$\tau\tau~(71.0\%)$ &	$hh~(18.4\%)$ & $b\bar{b}~(10.6\%)$ & no (DD)\\
BP2		&	$10.3$	&	$\tau\tau~(66.5\%)$ &	$hh~(20.8\%)$ & $b\bar{b}~(12.6\%)$ & no (DD)\\
BP3		&	$29.6$	&	$\tau\tau~(92.1\%)$ &	$b\bar{b}~(7.8\%)$ &  & yes \\ 
BP1		&	$56.8$	&	$b\bar{b}~(67.7\%)$ &	$Zh~(26.9\%)$ & $\tau\tau~(5.2\%)$ & no (DD)\\
BP2		&	$56.8$	&	$b\bar{b}~(72.9\%)$ &	$Zh~(21.2\%)$ & $\tau\tau~(5.6\%)$ & no (DD)\\
BP3		&	$59.4$	&	$b\bar{b}~(95.5\%)$ &       $\tau\tau~(3.5\%)$ &  & no (DD) \\
BP1		&	$59.9$	&	$b\bar{b}~(48.5\%)$ &       $Zh~(36.9\%)$ &  $\tau\tau~(14.4\%)$ &  no (DD, ID, CMB) \\
BP2		&	$59.9$	&	$b\bar{b}~(51.9\%)$ &       $Zh~(32.7\%)$ &  $\tau\tau~(15.2\%)$ &  no (DD, ID, CMB) \\
BP4		&	$60.6$	&	$b\bar{b}~(97.7\%)$ &       $\tau\tau~(1.7\%)$ & &   yes \\
BP3		&	$62.7$	&	$b\bar{b}~(78.4\%)$ &       $\tau\tau~(20.3\%)$ &  &   no (DD, ID, CMB) \\
BP4		&	$64.0$	&	$b\bar{b}~(79.8\%)$ &       $\tau\tau~(19.6\%)$ &  &   no (ID, CMB) \\
\botrule
\end{tabular}
\caption{Summary of all 16 benchmark points yielding the correct DM relic abundance, ordered by increasing light neutralino mass, $m_{\tilde{\chi}^0_1}$. For each point we give the dominant pair annihilation final states at zero temperature and their relative size (in $\%$). In order to summarize the findings in Fig.~\ref{fig:DM_Mneut1} we also indicate whether the benchmark points are allowed by the CMB, DM indirect detection (ID) and DM direct detection (DD) constraints, and if excluded, indicate the relevant constraint(s).}
\label{Tab:inddir}
\end{table*}

In summary, we find that 
\begin{itemize}
\item In the only unexcluded window at very light Higgs masses, $M_h\simeq 1\gev$, (cf.~BP1) generically three possible sub-$\mathrm{GeV}$ neutralino masses produce the right thermal relic density; the mass choice corresponding to $\mneut{1}\simeq M_h/2$ is excluded by limits from CMB distortions, while the other two viable mass values at $\mneut{1}\simeq 0.4\gev$ and $\mneut{1}\simeq 0.8\gev$ escape all current dark matter searches. Future direct detection experiments will however test this very low mass scenarios;
\item For Higgs masses just above the $\Upsilon(1S)$ mass threshold (cf.~BP2) four to five values of the lightest neutralino mass are compatible with the thermal relic dark matter density requirement; however, masses at and above $\mneut{1}\simeq M_h$ are robustly excluded by direct dark matter searches, leaving only a relatively narrow window of possible dark matter particle masses in the range between $\mneut{1}\lesssim M_{\Upsilon(1S)}/2$ and about 7 GeV. Again, we predict that all of these neutralino mass choices can be tested with future direct detection experiments; 
\item For intermediate light Higgs masses, $M_h$, between around $55\gev$ and $65\gev$ (cf.~BP3) the two possible neutralino masses correspond to $\mneut{1}\simeq M_h/2$, and yield direct detection rates very close to current LUX limits;
\item Finally, for $M_h\gtrsim 60$ GeV we find that the only possibility is resonant annihilation via quasi-on-shell pseudoscalar Higgs boson exchange with $\mneut{1}\lesssim M_A/2$, while the $\mneut{1}\gtrsim M_A/2$ side of the resonance is ruled out by indirect detection constraints.
\end{itemize}
It is remarkable that the light dark matter scenarios we study here will all be testable with future direct dark matter searches with an improvement of less than an order of magnitude compared to current constraints for $\mneut{1}\gtrsim 5$ GeV, and with the next generation low-dark matter mass experiments for lighter masses. Indirect detection constraints can also test the heavier neutralino scenarios, $\mneut{1}\gtrsim 50\gev$, with an improvement of around two orders of magnitude over current limits.

\section{Discussion and Conclusions}\label{sec:conclusions}
In this study we have addressed the possibility of light neutralino dark matter in the $\cp$-conserving MSSM, focusing on scenarios where the observed $125\gev$ Higgs state is the {\em heavier} of the two $\cp$-even MSSM Higgs bosons. This interpretation can be successfully realized in the limit of {\em alignment without decoupling}. In this context, since the $\mu$ parameter needs to be in the multi-$\mathrm{TeV}$ range, the lightest neutralino is gaugino-like, and, if light enough for sfermion coannihilation to be irrelevant, and for its wino content to be small, it must pair-annihilate through Higgs boson exchange to avoid over-closing the Universe with thermal relics.

Requiring a thermal relic neutralino density matching the observed cold dark matter density in the Universe, and taking into account all current constraints on direct and indirect dark matter searches, few selected possibilities exist for light neutralino dark matter pair-annihilating via a light $\cp$-even Higgs portal:
\begin{enumerate}
\item A {\em very light} Higgs and dark matter mass scenario, with the Higgs mass in a special mass range around 1 GeV which is not ruled out by meson decay constraints, and with a lightest neutralino mass in the vicinity of either $0.4\gev$ or $0.8\gev$ (cf.~BP1);
\item A {\em light} to {\em intermediate} Higgs and dark matter scenario, with the Higgs mass, $M_h$, ranging from just above the $\Upsilon(1S)$ mass threshold, $M_{\Upsilon} \sim 9.5\gev$, to around $60\gev$, and a neutralino mass $\mneut{1}\simeq M_h/2$ (cf.~BP2 and BP3);
\item A ``{\em heavy}'' dark matter scenario, with $M_A$ around or slightly larger than the heavier $\cp$-even Higgs $M_H\simeq 125\gev$, and $\mneut{1}\simeq M_A/2$ (cf.~BP4).
\end{enumerate}
In all four cases the expected direct detection rates are large enough to be within one order of magnitude of current limits, with the exception of the very light Higgs boson and dark matter mass scenario. For the latter, however, the predicted large direct detection cross sections are well within the experimental capabilities of planned, dedicated low mass dark matter search experiments such as \texttt{SuperCDMS}, \texttt{NEWS} and \texttt{CRESST-III}. 

Besides the promising prospects for dark matter direct detection experiments, the scenario under consideration here features a rich and interesting Higgs boson phenomenology that warrants a dedicated search program at the LHC. If a light $\cp$-even Higgs boson with mass $\lesssim 60\gev$ exists, the novel Higgs decay signatures $H^+\to W^+h$ and $A\to Zh$ appear, typically with rather sizable branching fractions. In particular, in contrast to the decay mode $H\to hh$, these decays are unsuppressed in the limit of alignment without decoupling, and thus offer a promising experimental avenue that complements the direct dark matter searches and Higgs rate measurements.

In conclusion, current experimental data on the Higgs sector allow for the possibility that the $125\gev$ Higgs state is the {\em heavier} of the two $\cp$-even Higgs bosons of the $\cp$-conserving MSSM. If this is the case, we showed here that light supersymmetric dark matter pair-annihilating via quasi-on-shell $s$-channel exchange of the {\em lighter} $\cp$-even Higgs boson is a generic possibility. The scenario is rather tightly constrained by both indirect and direct searches for dark matter, but several open windows exist, with possible neutralino masses ranging from a fraction of a GeV all the way up to around $65\gev$.

We have explored in detail all possible combinations of lightest neutralino and light Higgs masses compatible with thermal relic dark matter requirements and with current constraints from indirect and direct dark matter searches. Our results indicate that this \emph{light Higgs--light neutralino} scenario is both phenomenologically viable at present and experimentally testable in the near future, especially by forthcoming direct dark matter search experiments. Moreover, by demonstrating several successful light dark matter realizations of this scenario our study corroborates the necessity of specifically targeting this possible realization of the MSSM with dedicated searches at the LHC.

\section*{Acknowledgements}
We thank Howard Haber for helpful discussions.
This work is supported by the US Department of Energy, Contract DE-SC0010107-001. TS is also supported by a Feodor-Lynen research fellowship sponsored by the Alexander von Humboldt-foundation.

\bibliographystyle{apsrev4-1}
\bibliography{main}

\end{document}